\documentstyle[prl,aps,floats,epsf]{revtex}
\begin{document}
\draft

\twocolumn[\hsize\textwidth\columnwidth\hsize\csname
@twocolumnfalse\endcsname

\title{Charge dynamics and  metal-insulator transition 
in Si$_{1-x}$Gd$_{x}$ and Si$_{1-x}$Y$_{x}$  alloys.}
\author{D.N.~Basov$^{1}$, A.M.~Bratkovsky$^{2}$, 
P.F.~Henning$^{1}$, B.~Zink$^{1}$, F.~Hellman$^{1}$, Y.J.~Wang$^{3}$, 
C.C.~Homes$^{4}$, and M.~Strongin$^{4}$} 

\address{$^{1}$Department of Physics, 
University of California, San Diego, La Jolla, CA 92093-0319} 
\address{$^{2}$Hewlett-Packard Laboratories, 1501 Page Mill Road,  Palo
Alto, CA 94304} 
\address{$^3$ NHMFL, Florida State University, Tallahassee, FL 32310 } 
\address{$^{4}$Department of Physics, Brookhaven National 
Laboratory, Upton, NY 11973-5000} 
\date{May 30, 2000 } 
\maketitle 

\begin{abstract}

Carrier dynamics in amorphous a-Si$_{1-x}$RE$_{x}$ (RE=Gd, Y) films  has been 
studied in the doping regime close to the metal-insulator transition by 
means of infrared spectroscopy. Optical constants throughout the entire 
intra-gap region ($\hbar \omega <$ 1~eV) have been found to be anomalously 
sensitive to changes of temperature and/or magnetic field.  The observed 
behavior is consistent with the model of hopping transport where the 
interaction of carriers with  both the lattice and large core spin of Gd 
ions is taken into account. 
\pacs{71.30.+h, 71.38.Ht, 72.20.Ee, 78.20.Ls, 78.66.Jg} 
\end{abstract} \vskip 2pc ] % end \twocolumn[...] 

Magnetic dopants may radically alter the properties of prototype
semiconductors, such as Si or GaAs, and may lead to extraordinary 
effects even in these fairly simple systems\cite
{ohno,samarath,hellman96}. For example, III-V materials doped with Mn
develop ferromagnetism with the Curie temperature exceeding 110 K\cite{ohno}
and amorphous Si alloyed with a high concentration of Gd shows gigantic
negative magnetoresistance\cite{hellman96}.  Present understanding of
the role of magnetic impurities in carriers dynamics, especially in the 
regime where a material is close to the boundary of the metal-insulator 
transition (MIT), is rather poor, in part because the experimental results 
are very scarce. This motivated us to perform studies of the electromagnetic 
response of a-Si$_{1-x}$Gd$_{x}$  in the vicinity of the MIT 
using the techniques of infrared (IR) optics and magneto-optics,
which are ideally suited for the task. 

We have measured the IR spectra of heavily doped a-Si$_{1-x}$Gd$_{x}$
and its non-magnetic counterpart a-Si$_{1-x}$Y$_{x}$ close to the
boundary of the metal-insulator transition. The amorphous Si 
matrix allows one to increase the concentration of magnetic Gd ions beyond 
the solubility limit in crystalline silicon. At $x\simeq 0.11-0.13$ these 
materials are insulating at  $T\rightarrow 0$ in zero magnetic field but 
become conducting in the field $H\simeq $ 5T\cite{hellman96}. The effect is 
reminiscent of the behavior of diluted magnetic 
semiconductors\cite{tokura}. However, in the latter case negative 
magnetoresistance (MR) is usually observed close to a temperature of 
magnetic ordering\cite{tokura}. On the contrary, a-Si$_{1-x}$Gd$_{x}$ 
alloys show large negative MR at all temperatures below $\simeq 80$K, and 
the magnitude of the MR is not affected by spin-glass-like freezing at 
$T<T_{f}=5-10$K. It is worth mentioning that the standard theory of 
electron hopping predicts {\it positive} MR in amorphous semiconductors 
\cite{efros} whereas a-Si$_{1-x}$ Gd$_{x}$ exhibits huge {\em negative} MR. 
We have found that one particular signature of the MIT in these systems is 
associated with the massive transfer of the spectral weight over an energy 
interval exceeding 1~eV. This scale is  larger than the 
typical magnitudes of the Coulomb gap or the correlation gaps forming at 
the chemical potential in the density of states in the insulating samples 
\cite{efros,hertel}. At the same time the interaction of carriers with the 
lattice is known to be particularly strong in amorphous 
solids\cite{pwa75,elliott78} and, together with exchange coupling of 
carriers to Gd local spins, it allows us to account for the  
optical properties of a-Si$_{1-x}$ Gd$_{x}$. 

The transmission coefficient $T(\omega )$ has been measured for $\simeq $
0.5 $\mu $m a-Si$_{1-x}$Gd$_{x}$ films prepared by electron beam
co-evaporation\cite{hellman96} on a high-resistance ($\sim 10$k$\Omega $cm)
Si substrate. Here we report the data for the ``critical'' sample with
12~at.$\%$Gd, 
in which the dc conductivity $\sigma _{{\rm dc}}$ vanishes at
$T\rightarrow 0$ and of a 
``metallic'' sample with 15~at.$\%$ Gd in which $\sigma _{\rm
dc}(T\rightarrow 0)\simeq 80$~$\Omega ^{-1}$cm$^{-1}$ increases up to
300$~
\Omega ^{-1}$cm$^{-1}$ at $T=300$K. In order to isolate the effects
associated with magnetic interactions, we compare the Gd-doped samples (core
f-spin $S=7/2)$ with films containing similar amount of non-magnetic Y. The $%
T(\omega )$ spectra were measured over the frequency interval 20-10000~cm$%
^{-1}$. We have obtained the absolute values of $T(\omega )$ in zero field
and then measured changes of the transmission at 10 K induced by magnetic
field up to 15 T. Complex conductivity $\sigma _{1}(\omega )+i\sigma
_{2}(\omega )$ was determined from the Kramers-Kronig analysis by taking
into account the optical constants of the substrate.

The dominant feature of the $\sigma_1(\omega)$ spectra of all films is
an absorption edge around 10,000 cm$^{-1}$, Fig.~1. This feature can
be attributed to the gap 
in the density of states of amorphous Si. We have also found significant
absorption throughout the intragap region that is dramatically affected by
changes in temperature and magnetic field. To quantify the strength of the
intragap absorption, we define the spectral weight \cite{weight} 
\begin{equation}
N_{eff}(\omega )=\int_{0}^{\omega }d\omega ^{\prime }\sigma _{1}(\omega
^{\prime }).  \label{eq:neff}
\end{equation}
The magnitude of $N_{eff}$ is proportional to $n/m^{\ast }$ where $n$ is the
density of carriers participating in the conductivity at energies below $%
\omega $ and $m^{\ast }$ is their mass. The integration up to $\omega =$
5,000 cm$^{-1}$\cite{cut} gives the following estimates of the carrier
density, under the assumption that $m^{\ast }=m_{e}$: $n=3.9\times 10^{20}$
cm$^{-3}$ for the critical sample and $n=5.4\times 10^{20}$ cm$^{-3}$ for
metallic samples.

All films reveal a broad resonance at $\omega \simeq 700-1000$ cm$^{-1}$; the resonance becomes more pronounced with
increasing concentration of dopants. This feature is characteristic of
hopping conductivity in disordered semiconductors, as will be discussed
below. The absolute value of the $\sigma_1(\omega)$ in the far IR region is
strongly suppressed at low $T$\cite{grunner}. Extrapolation of the 
conductivity to $\omega \rightarrow 0$ results in good agreement with the $%
\sigma _{{\rm dc}}$ values. Because of this suppression of the conductivity,
part of the far IR spectral weight appears to be lost. Integration of the $%
\sigma _{1}(\omega )$ spectra shows that the missing weight is not recovered
at $\omega <1$ eV. According to the f-sum rule, $N_{eff}(\infty )=n/m_{e}=%
{\rm const},$ so the total area under the $\sigma _{1}(\omega )$ spectrum
must remain constant. We conclude, therefore, that the spectral weight
disappearing from the intragap region is transferred to frequencies above
the gap value. This result is anomalous since changes of temperature by
200-300K have a conspicuous impact on the spectra of $\sigma
_{1}(\omega )$ over the energy interval exceeding 1~eV (12000 K). We
are unaware of an accurate analysis
of the oscillator strength throughout the intragap region in other amorphous
semiconductors. Unusual transfer of the spectral weight is known to occur in
systems with strong correlations between the electrons\cite{tokura,thomas}.
We emphasize that at $T>80$K both Y- and Gd-doped samples show similar
behavior suggesting that these effects are common in both magnetic and
non-magnetic alloys.

The decline in the conductivity of a-Si$_{1-x}$Gd$_{x}$ samples continues at
low temperatures $\lesssim 70-80$K; the conductivity of a-Si$_{1-x}$ Y$%
_{x}$ films is only weakly temperature dependent. This additional reduction
of the conductivity drives the critical Gd-doped sample towards the
insulating state, whereas a-Si$_{1-x}$Y$_{x}$ film with similar doping
remains ``metallic'': $\sigma _{{\rm dc}}(T\rightarrow 0)\approx 90 
\Omega^{-1}$cm$^{-1}
$. The spectral weight, lost from the far IR region at $T<80$K in 
a-Si$_{1-x}$Gd$_{x},$ is also transferred to energies beyond 1 eV.

Changes in the conductivity of a-Si$_{1-x}$Gd$_{x}$ films induced by
magnetic field are even more anomalous. In applied
field both metallic and critical films recover most of the spectral weight,
which was lost from the conductivity at $T<80$ K. Field-induced enhancement
of $N_{eff}$ is directly related to the large negative MR found in Gd-based
films. Remarkably, the magneto-conductivity $\Delta \sigma _{H}=\sigma
_{1}(\omega ,H)-\sigma _{1}(\omega ,H=0)$ of both ``critical'' and
``metallic'' films remains positive up to at least 0.5 eV. This
frequency region exceeds the energy scale associated with the magnetic
field by more than 2 orders of magnitude. IR 
magnetoconductivity of Y-based films is negligibly small.

Insets in Fig.~1 show the $\Delta \sigma _{H}$ and $\Delta \sigma
_{T}=\sigma _{1}(\omega ,T)-\sigma _{1}(\omega ,$10K$)$ spectra indicating
important distinctions in the response of magnetic and non-magnetic films.
The differential spectra show monotonic (``Drude''-like) dependence in
Y-based samples and in the metallic Gd-based samples. In contrast, both $%
\Delta \sigma _{T}$ and $\Delta \sigma _{H}$ for the critical a-Si$_{1-x}$Gd$%
_{x}$ film display a peak centered at around 10 meV. This latter behavior is
characteristic of disordered semiconductors, where carriers are localized,
with the peak position corresponding to the magnitude of the mobility gap 
\cite{allen}. The 10 meV feature may account for the enhancement of
localization in magnetic Gd-based films. Interestingly, the energy
associated with the peak agrees quantitatively with the temperature at
which the localization effects become prominent in the transport
measurements. Indeed, the dc conductivities of a-Si$_{1-x}$Gd$_{x}$ and a-Si$%
_{1-x}$Y$_{x}$ films for $x\simeq 0.11-0.13$ are identical at $T>80$ K.
However, below this temperature $\sigma _{{\rm dc}}(T)$ in a-Si$_{1-x}$Gd$%
_{x}$ is suppressed faster than that of Y-based film and eventually vanishes
at $T\rightarrow 0$. On the contrary, the Y-based samples remain ``metallic''
\cite{hellman96}. It is reasonable to ascribe the difference between a-Si$%
_{1-x}$Gd$_{x}$ and a-Si$_{1-x}$Y$_{x}$ films to magnetic exchange
interactions, which are present only in the Gd-based system. We can estimate
the exchange coupling between the core $f$- and the carrier spin as $JS\sim $%
10 meV\cite{JSGd}, which is in good agreement with localization features
seen both in transport and IR experiments. We do not observe the
corresponding peak in the spectra for {\em metallic} Gd-based films. The
localization trends are much less pronounced in this sample, and one can
speculate that the mobility gap is shifted there to lower energies, so that
the signature of this effect in $\sigma _{1}(\omega )$ occurs below our low-$%
\omega $ cutoff.

We emphasize that the energy scale involved in transfer of the spectral
weight with decreasing $T$ and partial recovery of $N_{eff}$ with increasing
field $H$ extends beyond 1 eV. Therefore, this scale dramatically exceeds
the mobility gap in the studied films as well as the Zeeman energy for a- Si$%
_{1-x}$Gd$_{x}$ by at least  two orders of magnitude. Looking for possible
origins of the effects, extending over such unexpectedly large energy scale,
one can recall that the shift of the electronic levels to lower energies due
to interaction with the lattice (so-called polaron shift $E_{p}$) can
readily reach 1 eV in many semiconductors\cite{abopt99}. A strong
electron-lattice interaction is favored by the ``softness'' of an amorphous
matrix and it leads to formation of lattice polarons and local singlet pairs
(Anderson bipolarons) \cite{pwa75,elliott78}. To illustrate a possible
effect of lattice (bi)polarons we consider the following model. We assume
that the energies of localized carriers at site ${\bf i}$ and spin $%
s=\uparrow \left( \downarrow \right) $ are split by the external field $H$
and the exchange interaction with the core f-spins $S$ ($S=7/2$ for Gd and $0
$ for Y) 
\begin{equation}
\epsilon _{{\bf i\uparrow }\left( \downarrow \right) }=\epsilon _{{\bf i}%
}^{0}-(+)\frac{1}{2}JS\eta -(+)\mu _{B}H  \label{eq:engy}
\end{equation}
where $\eta =\left\langle S_{z}\right\rangle /S$ is the relative f-spin
polarization, $JS\simeq 10$ meV as discussed above, $\mu _{B}$ is the Bohr
magneton, the impurity levels $\epsilon _{{\bf i}}^{0}$ are distributed with
the width $\Gamma $ $\left( \sim 0.2\text{eV \cite{castilho89}}\right) $.
This splitting results in the polarization of the carriers (lattice
polarons) $\mu =n_{p\uparrow }-n_{p\downarrow },$ where $n_{ps}$ is the
density of polarons with spin $s=\uparrow \left( \downarrow \right) $.
Within this picture, the pairs constitute the ground state, with the
individual polarons existing as thermal excitations\cite{pwa75,elliott78}.
The electron-phonon coupling in amorphous semiconductors is known to be
sufficiently strong to bind carriers into singlet spin pairs with the
binding energy $\Delta \sim 30$meV, given by the difference between
electron-phonon pairing and Coulomb on-site repulsion\cite{pwa75}. An
estimate for $\Delta $ in the studied systems can be found from the
magnitude of the ``hard'' gap seen in tunneling data\cite{hertel}.

Both single lattice polarons and bound pairs produce the Gaussian-like
absorption resonance $\sigma _{1}(\omega )$\cite{abopt99,bryksin83,mahan}.
The conductivity due to excitation of single polarons is peaked at $\omega
_{p},$ which is proportional to $E_{p},$ and has the form $\sigma
_{p}(\omega )\propto \lbrack n_{ps}/\omega \gamma ]\exp $ $\left[ -(\omega
-\omega _{p})^{2}/\gamma ^{2}\right] $, where $\gamma (\approx 30$meV) is
the width of the absorption resonance. Analogously, the conductivity
of the bound pairs is described as $\sigma _{bp}(\omega )\propto \lbrack
n_{bp}/\omega \gamma ]\exp \left[ -(\omega -\omega _{bp})^{2}/\gamma ^{2}%
\right] $, where $n_{bp}$ is their density and $\omega _{bp}=\omega
_{p}+\Delta ,$ with $\Delta $ being the binding energy of the pairs \cite
{bryksin83,mahan,abopt99}. The scatter of the energies of the impurity
states, $\Gamma ,$ results in observed asymmetric shape of the absorption
line\cite{bryksin83}. The hopping conductivity is the sum of both
contributions: $\sigma _{{\rm hopping}}=\sigma _{p}+\sigma _{bp}$. In order
to calculate the spectra presented in Fig.~2 we add an interband
contribution $\sigma _{\text{{\rm inter}}}$ to $\sigma _{{\rm hopping}}$.
The standard form for $\sigma _{{\rm inter}}$ in amorphous semiconductors is 
$\sigma _{{\rm inter}}\propto (\omega -\omega _{0})^{2}/\omega $, with the
threshold energy $\omega _{0}\approx 0$\cite{mottdavis}. The calculated
conductivity $\sigma _{{\rm hopping}}(\omega )+\sigma _{{\rm inter}}(\omega )
$ at different $T$ (Fig.~2, left panel of ) is in fair agreement with the
experimental data. In particular, the model accounts for massive transfer of
the spectral weight observed in our films at low temperatures. This is due
to the increase of the density of single lattice polarons $n_{p}\propto \exp %
\left[ -(\Delta -JS\eta -2\mu _{B}H)/2k_{B}T\right] $ with temperature as
the result of thermal dissociation of the local spin pairs. An important
feature of this model is that a fraction of the spectral weight associated
with the higher order hopping processes appears at energies comparable to
the polaron shift. Although the detailed account for these processes is
beyond the saddle-point approximation used to estimate $\sigma _{{\rm hopping%
}}$ \cite{mahan,bryksin83} our model shows that at higher $T$ their
contribution is reduced, leading to an enhancement of the low-energy
spectral weight.

Another feature of the polaronic scenario is that it results in negative
optical magnetoresistivity observed in a-Si$_{1-x}$Gd$_{x}$, provided that
the exchange coupling of carriers with a large core f-spin $S$ of Gd is
taken into account. According to Eq.~(\ref{eq:engy}), lattice polarons are
polarized by the external field $H$ {\em and} by much stronger molecular
field $JS\eta /\mu _{B}$. Their polarization $\mu $ produces the exchange
field $J\mu /\left( 2g_{\text{{\rm Gd}}}\mu _{B}\right) $ which in return
polarizes the f-spins $S_{z}$. The polarizations $\eta $ and $\mu $ are
determined self-consistently, analogous to Ref.~\cite{abopt99}. The exchange
interaction strongly enhances the polarization of carriers $\mu $ in the
external field. Since the spin of lattice polarons interacts with the
exchange and external fields, the polaron conductivity of Gd-doped films $%
\sigma _{p}(\omega )$ acquires a field dependent factor, $\sigma _{p}(\omega
,H)=\sigma _{p}(\omega ,0)\cosh \left[ \left( \frac{1}{2}JS\eta +\mu
_{B}H\right) /k_{B}T\right] $ \cite{magpolaron}. This term increases rapidly
with the applied field, as does the density of polarons. The net results are
displayed in  Fig.~2. One finds a strong enhancement of
the $\sigma _{p}(\omega )$ contribution to the total hopping conductivity
and very large negative MR due to the transfer of the spectral weight to
lower frequencies. While some models were suggested for the spectral weight
transfer with the onset of ferromagnetism \cite{abcmr,tokura}, the novelty
of the proposed polaronic scenario is that the effect occurs in the {\em %
paramagnetic} state. Thus, the polaronic model introduced above not only
reproduces the principal features of the conductivity spectra but also
accounts for the non-trivial behavior of the spectral weight at finite
temperature and/or field. This agreement highlights the prominent role of
the electron-lattice interaction in the correlated behavior of doped
amorphous silicon\cite{magpolaron}.

Important open issue concerns a possible coexistence of extended and
localized states on the metallic side of the transition\cite{dor}. Indeed,
the frequency dependence of the $\sigma _{1}(\omega ,T=0)$ spectra of the
``metallic'' samples is similar to what is observed at finite $T$ or $H$ in
the optical response of the critical samples. As shown above, these latter
regimes are consistent with the co-existence of both types of states at the
chemical potential, indicating that similar co-existence may occur at $T=H=0$
on the ``metallic'' side of the transition. We also note that the proposed
polaronic scenario requires the presence of nonbound carriers at $%
T\rightarrow 0$ even in the critical films in order to explain negative MR
of a-Si$_{1-x}$Gd$_{x}$ at very low temperatures\cite{hellman96}. This
requirement stems from the fact that the magnetoresistance is related
to the
existence of unpaired polarons, since only they are subject to exchange
interaction with the core spins on the Gd sites. We conclude by noting that
properties of a-Si$_{1-x}$Gd$_{x}$ alloys studied in the present work with
the use of IR spectroscopy are likely to be generic to other classes of
amorphous semiconductors with magnetic impurities. Work at UCSD is supported
by NSF and DOE.

\newpage 
\begin{figure}[tbp]
\caption{ The optical conductivity of a-Si$_{1-x}$RE$_{x}$ 
films at different temperatures (middle and bottom panels) 
and magnetic fields (top panels). Insets show differential conductivity 
$\Delta \protect\sigma_T( \protect\omega) $ and 
$\Delta\protect\sigma_H(\protect\omega)$ defined in the text. In all 
samples the intragap conductivity is strongly reduced at low $T$ with the 
spectral weight being transferred to energies exceeding the band gap of 
amorphous Si. In a-Si$_{1-x}$Gd$_{x}$ films a part of the spectral weight, 
which is lost at $T<80$ K, is recovered in high magnetic field (top 
panels).} \label{fig:fig1} \end{figure} 

\newpage 
\begin{figure}[tbp]
\caption{ The calculated optical conductivity $\protect\sigma (\protect%
\omega; T,H)$ of a-Si$_{1-x}$Gd$_{x}$ for $x=0.12$ at different temperatures
and magnetic fields. Gray lines in the left panel show separate
contributions to the conductivity associated with the polarons (dashed line)
and spin pairs (solid line) at $T=300$K. The
polaron absorption peak is at $\protect\omega_{p}=200$ cm$^{-1}$. The 
insets show $\Delta\protect \sigma_T(\protect\omega)$ and 
$\Delta\protect\sigma_H(\protect\omega)$. The model spectra reproduce the 
key features of the experimental $\protect\sigma (\protect \omega )$ 
presented in Fig.~1, as well as the redistribution of the spectral weight 
induced by temperature (left panel) and magnetic field (right panel).} 
\label{fig:fig2} \end{figure}


\begin{references}
\bibitem{weight}  $N_{eff}(\infty )$ would be equal to an atomic density of
carriers with an extra factor $2m_{e}\Omega _{c}/\pi e^{2}$, where $m_{e}$
is the bare electron mass and $\Omega _{c}$ is the atomic volume.
\bibitem{cut}  The choice of $\omega _{c}=5000$ cm$^{-1}$ is suggested by
the frequency dependence of the $\sigma _{1}(\omega )$ spectra shown in
Fig.~1. At $\omega >5000$ cm$^{-1}$ the conductivity is clearly dominated by
transitions across the band gap of Si with significantly broadened band
tails in the present amorphous samples.
\bibitem{magpolaron}  
We consider the lattice polarons only. The ``magnetic polaron''
(carrier, localized as a result of its interaction with magnetic
background) is unstable because of the high carrier density in our
films ($\gtrsim 10^{21}$ cm$^{-3}$).
\end{references}
\end{document}